\documentclass[11pt,leqno,twoside]{article}
\usepackage{geometry}
\usepackage{amsmath}
\usepackage{amsfonts}
\usepackage{amsthm}
\usepackage{indentfirst}

\numberwithin{equation}{section}

\theoremstyle{plain}
\newtheorem{theorem}{\indent\rm T\,h\,e\,o\,r\,e\,m\;}[section]

\newtheorem{proposition}[theorem]{\indent\rm P\,r\,o\,p\,o\,s\,i\,t\,i\,o\,n\;}

\theoremstyle{definition}

\theoremstyle{remark}
\newtheorem{remark}[theorem]{\indent\rm R\,e\,m\,a\,r\,k\;}

\makeatletter                                                  %
\renewcommand*{\@seccntformat}[1]{
  \csname the#1\endcsname\;-                                   %
}                                                              %
\renewcommand{\section}{\@startsection{section}{1}{0mm}        %
   {1.5\baselineskip}
   {1\baselineskip}
   {\indent\normalfont\normalsize\bfseries}
   }                                                           %
\renewcommand*{\@seccntformat}[1]{
  \normalfont\bfseries\csname the#1\endcsname\;-               %
}                                                              %
\renewcommand\subsection{\@startsection                        %
  {subsection}{2}{0mm}
  {1.5\baselineskip}
  {1\baselineskip}
  {\indent\normalfont\normalsize\itshape}}
\renewcommand*{\@seccntformat}[1]{
  \normalfont\bfseries\csname the#1\endcsname\;-               %
}                                                              %
\renewcommand\subsubsection{\@startsection                     %
  {subsubsection}{2}{0mm}
  {1.5\baselineskip}
  {1\baselineskip}
  {\indent\normalfont\normalsize\texttt}}
\makeatother                                                   %
\newcommand		{\N}		{\mathbb N}			
\newcommand		{\R}		{\mathbb R}			

\renewcommand	{\L}		{\mathcal L}		%

\newcommand		{\lt}			{\left}				%
\newcommand		{\rt}			{\right}			%
\renewcommand	{\(}			{\lt(}
\renewcommand	{\)}			{\rt)}

\newcommand		{\com}[1]		{\lt[{#1}\rt]}		

\newcommand		{\nrm}[1]		{\lt\lVert #1\rt\rVert}
\newcommand		{\Nrm}[2]		{\nrm{#1}_{#2}}

\DeclareMathOperator{\tr}		{Tr}				
\DeclareMathOperator{\diag}		{diag}

\newcommand		{\jj}			{\mathrm{j}}	
\newcommand		{\init}			{\mathrm{in}}
\newcommand		{\fb}			{\mathfrak b}
\newcommand		{\eps}			{\varepsilon}

\usepackage{braket}

\newcommand		{\op}		{\boldsymbol{\rho}}	

\newcommand		{\opp}		{\boldsymbol{p}}

\newcommand		{\Dh}		{\boldsymbol{\nabla}}	
\newcommand		{\Dhx}		{\Dh_{\!x}}				
\newcommand		{\Dhv}		{\Dh_{\!v}}			
\newcommand		{\Dhxj}		{\Dh_{\!x_\jj}}			
\newcommand		{\Dhvj}		{\Dh_{\!v_\jj}}		

\begin{document}
\thispagestyle{empty}


\centerline{\large{\textsc{Weakly interacting Fermions:}}} 

\centerline{\large{\textsc{mean-field and semiclassical regimes}}}

\begin{center}
{\sc Chiara Saffirio} \
\end{center}

\renewcommand{\thefootnote}{\fnsymbol{footnote}}

\footnotetext{
This research was supported by the Swiss National Science Foundation through the Eccellenza project PCEFP2\_181153 and the NCCR SwissMAP.}

\renewcommand{\thefootnote}{\arabic{footnote}}
\setcounter{footnote}{0}

\vspace{0.6cm}
\begin{center}
\begin{minipage}[t]{11cm}
\small{
\noindent \textbf{Abstract.}
The derivation of effective macroscopic theories approximating microscopic systems of interacting particles is a major question in non-equilibrium statistical mechanics. In these notes we present an approximation of systems made by many fermions interacting via inverse power law potentials in the mean-field and semiclassical regimes, reviewing the material presented at the 11th summer school ``Methods and Models of Kinetic Theory'' held in Pesaro in June 2022.\\ 
More precisely, we focus on weakly interacting fermions whose collective effect can be approximated by an averaged potential in convolution form, and review recent mean-field techniques based on second quantization approaches. As a first step we obtain a reduced description given by the time-dependent Hartree-Fock equation. As a second step we look at longer time scales where a semiclassical description starts to be relevant and approximate the many-body dynamics with the Vlasov equation, which describes the evolution of the effective probability density of particles on the one particle phase space.  
\medskip

\medskip

\noindent \textbf{Mathematics~Subject~Classification:}
82C10, 35Q41, 35Q55, 82C05, 35Q83.

}
\end{minipage}
\end{center}

\bigskip

\section{Introduction}

Complex physical systems made of a large number $N$ of components are a central topic in modern science. The microscopic description of these systems is given in terms of the elementary components and their interactions, resulting in complicated behaviour. Although very accurate, the microscopic description is not well-suited for computations due to the high number of degrees of freedom. Thus it is useful to look at the systems at different (macroscopic) scales, where the description focuses only on macroscopic observable quantities that retain systems' collective behaviour, statistical properties and effective interactions. This picture, often called effective theory, is certainly less accurate than the microscopic one, but more suitable from a computational viewpoint. For this reason it is important to understand the emergence of the macroscopic effective theories from the microscopic description. A natural way of doing it is by means of scaling limits.  \\
In these notes we focus on the description of gases made by a large number ($10^3-10^{23}$) of interacting particles and look for approximations given at the kinetic scale, that is on quantities which averages are susceptible of measurement. \\
 More specifically, consider a system of $N$ interacting quantum particles in $\mathbb{R}^3$. A state of the system is described by the wave function $\psi_N\in L^2(\mathbb{R}^{3N})$ with $\|\psi_N\|_{L^2}=1$, $|\psi_N(x_1,\dots,x_N)|^2$ representing the probability of finding the $N$ particles at $(x_1,\dots,x_N)$, with $x_i\in\mathbb{R}^3$ for $i=1,\dots, N$. An observable of the quantum system is a self-adjoint operator on $L^2(\mathbb{R}^{3N})$. If $A$ is a self-adjoint operator on $L^2(\mathbb{R}^{3N})$, its expectation in the state $\psi_N$ is given by the inner product $\langle \psi_N,A\,\psi_N\rangle$. For example, if $A$ is the identity operator $\mathbf{1}$, its expectation in the state $\psi_N$ is the $L^2$ norm of $\psi_N$. \\
We distinguish between two kind of quantum particles: bosons, whose wave function is symmetric in the exchange of particles, i.e.~$\psi_N\in L^2_{\rm s}(\mathbb{R}^{3N})$; and fermions, whose wave function is antisymmetric in the exchange of particles, i.e.~$\psi_N\in L^2_{\rm a}(\mathbb{R}^{3N})$.
The aim of these notes is to review the state of the art on the the time evolution of low energy fermionic states (at zero and positive temperature) in a mean-field limit coupled with a 
semi-classical regime.\\
{The paper is organised as follows: in section~\ref{sec:2} we describe the fermionic mean-field regime we are interested in and the class of relevant initial states, we introduce the Hartree-Fock and the Vlasov equations, we briefly present the state of the art and we conclude the section with the statement of our main result. In section~\ref{sec:MF} we present the main steps and techniques to prove the converge of the many-body dynamics to the Hartree-Fock equation. To this end we introduce the Fock space and the second quantization formalism. Section~\ref{sec:4} is devoted to the semiclassical approximation of the Hartree-Fock equation with the Vlasov dynamics. We then conclude with a brief account of open problems in the field, presented in section~\ref{sec:5}.}

\section{Mean-field and semiclassical regimes}\label{sec:2}

Consider a system of $N$ fermions interacting through a potential $K$ with associated Hamilton operator given by
\begin{equation*}
H_N^{\rm trap}=\sum_{j=1}^N-\frac{1}{2}\Delta_{x_j}+V^{\rm trap}(x_j)+\lambda\sum_{1\leq k<j\leq N} K(x_j-x_k)
\end{equation*}
where $V^{\rm trap}$ is an external potential confining the system in a volume of order one, and $\lambda$ is a coupling constant to be determined depending on the regime we are interested in. The mean-field regime corresponds to a choice of the parameters $\lambda$ and $N$ such that $\lambda$ is small, thus modelling the weak interaction among particles, and the number of particles $N$ is large. Although the interactions between couples of particles are weak, the total interaction should be of order one to survive in the limit of $N$ large. More precisely, for states $\psi_N$ confined in a volume of order one by the trapping external potential, the expectation of the kinetic and the potential energy are given by
\[
\langle \psi_N,\sum_{j=1}^N-\Delta_{x_j}\psi_N\rangle = O(N^{5/3})
\]
and 
\[
\langle \psi_N,\lambda \sum_{1\leq j<k\leq N}K(x_j-x_k)\psi_N\rangle = O(\lambda\,N)
\]
respectively. Notice that the bound on the kinetic energy follows from the Lieb-Thierring inequality 
\[
\langle \psi_N,\,\sum_{j=1}^N-\Delta_{x_j}\psi_N\rangle\geq C\int\rho_{\psi_N}(x)^{5/3}dx
\] 
where 
\[
\rho_{\psi_N}(x):=N\int |\psi_N(x,x_2,\dots,x_N)|^2dx_2\dots dx_N.
\]
Since we are interested in capturing the interacting behaviour of the system at macroscopic scale, we balance the kinetic and the potential energy by choosing $\lambda=N^{-1/3}$. To study the dynamics of the gas out of equilibrium we analyse how the system reacts to changes of the external potential. In particular, if we switch off the trapping potential $V^{\rm trap}$ we observe a non trivial time evolution given by the solution of the many-body Schr\"{o}dinger equation
\begin{equation}
i\partial_\tau\psi_{N,\tau}=\left[\sum_j-\frac{1}{2}\Delta_{x_j}+N^{-1/3}\sum_{1\leq j<k\leq N} K(x_j-x_k)\right]\psi_{N,\tau}.
\end{equation}
Observe that the kinetic energy per particle is of the order $O(N^{2/3})$, hence the average velocity per particle is of the order $O(N^{1/3})$. Notice moreover that the evolution of the position operator is proportional to the momentum operator. This is easy to see in the case of free (non-interacting) fermions:
\begin{equation*}
e^{-i\tau\Delta}\,x\,e^{i\tau\Delta}=x+2\,\tau\,i\,\nabla.
\end{equation*}
Thus if the typical velocity per particle is $O(N^{1/3})$ and system is in a volume of order $O(1)$, the time scale of the observer is of the order $O(N^{-1/3})$. This is a manifestation of a separation of scales between the wavelength and the length of the observables which is typical of the semiclassical regime. Indeed, rescaling the time variable as
\[
t=N^{1/3}\tau
\]  
(and therefore considering times of order one) 
yields
\begin{equation}\label{eq:scaled-MBS}
iN^{1/3}\partial_t\psi_{N,t}=\left[\sum_{j=1}^N-\frac{1}{2}\Delta_{x_j}+N^{-1/3}\sum_{1\leq j<k\leq N}K(x_j-x_k)\right]\psi_{N,t}.
\end{equation}
In the rest of the paper let 
\begin{equation}\label{eq:scaling}
\hbar=N^{-\frac{1}{3}}.
\end{equation}
By multiplying equation \eqref{eq:scaled-MBS} by $\hbar^2$ yields
\[\label{eq:MBS}
i\,\hbar\,\partial_t\psi_{N,t}=\left[\sum_{j=1}^N-\frac{\ \hbar^2}{2}\Delta_{x_j}+\frac{1}{N}\sum_{1\leq j<k\leq N}K(x_j-x_k)\right]\psi_{N,t}.
\] 
Notice that $\hbar$ plays the role of the reduced Panck constant, hence the fermionic mean-field limit ($N\gg 1$) is coupled to a semiclassical regime ($\hbar\ll 1$). In other words, we are looking at time and space length scales at which the Planck constant is small.

\subsection{Initial states}

The macroscopic behaviour of the system strongly depends on the assumptions on the initial state. In particular, we observe that if the initial state is uncorrelated and such  absence of correlations is preserved also at positive times $\psi_{N,t}$, a local averaging mechanism is expected to take place because of the strong law of large numbers, namely
\[
\langle \psi_{N,t}\,,\,\frac{1}{N}\sum_{1\leq j<k\leq N}K(x_j-x_k)\,\psi_{N,t} \rangle \sim \langle \psi_{N,t}\,,\,\sum_{j=1}^N (K*\varrho_t)(x_j)\,\psi_{N,t} \rangle
\]
where $\varrho_t(x_j)$ is the density of particles at $x_j$. \\
Due to the Pauli principle, fermionic uncorrelated states do not exist. Indeed, the Fermi statistics imposes that two particles cannot occupy the same quantum state, so that correlations are built in the statistics itself. However, it is well-known that the less correlated fermionic states are Slater determinants, i.e.~$N$-particle wave functions of the  form
\[\begin{split}\label{eq:Slater}
\psi_{N}^{\rm Slater}(x_1,\dots,x_N)&=\frac{1}{\sqrt{N!}}{\rm det}(f_j(x_k))_{1\,\leq\,j,\,k\,\leq\,N}\\
&=\frac{1}{\sqrt{N!}}\sum_{\pi}{\rm sign}(\pi)f_1(x_{\pi(1)})\dots f_N(x_{\pi(N)}),
\end{split}
\]
with $f_j\in L^2(\mathbb{R}^3)$ and $\{f_j\}_{j=1}^N$ orthonormal system. \\
The one-particle reduced density associated to a Slater determinant \eqref{eq:Slater} is the one-particle operator $\op_{N}^{(1)}$ with kernel
\[
\op_{N}^{(1)}(x,y)=\int_{\mathbb{R}^{3(N-1)}}\Psi_{N}^{\rm Slater}(x,x_2,\dots,x_N)\overline{\Psi_{N}^{\rm Slater}(y,x_2,\dots,x_N)}\,dx_2\dots dx_N\,,
\]
that can be rewritten, using \eqref{eq:Slater}, as orthogonal projection onto the space ${\rm span}\{f_1,\dots,f_N\}\subset L^2(\mathbb{R}^3)$, i.e.
\[\label{eq:Slater-1part}
\op_{N}^{(1)}=\sum_{j=1}^{N}\left|f_j\right\rangle\left\langle f_j\right|,
\]
using the bra-ket notation $|\cdot\rangle\langle\cdot|$. 
This class of one-particle density matrices is a good approximation for the static mean-field problem at zero temperature, namely one-particle density matrices of the form \eqref{eq:Slater-1part} minimize the Hartree-Fock energy functional.\\
In this paper we wish to consider more general states.
To this end, we recast the time evolution problem in terms of density matrices. Let $\boldsymbol{\rho}_N$ be a $N$-particle density matrix, i.e.~a self-adjoint operator acting on $L^2(\mathbb{R}^{3N})$. 
By the spectral theorem we have
\[
\op_{N}=\sum_{j\geq 1}\lambda_j|\psi_j\rangle\langle\psi_j|,
\]
with $\lambda_j\geq 0$ and $\{\psi_j\}_{j\geq 1}$ orthonormal system of antisymmetric wave functions $\psi_j\in L^2_a(\mathbb{R}^{3N})$. With an abuse of notation $\op_N(x_1,\dots,x_N,y_1,\dots,y_N)$ will denote the kernel of the operator $\op_N$.\\
The initial states we will focus on are called quasi-free states. These are states that are completely characterised by their one-particle reduced density matrix $$\op_{N}^{(1)}:=\tr_{2\dots N}\(\op_N\)$$ through the Wick rule {(see for instance \cite{Derezinski,Solovej})}, i.e.~the $k$-particle density matrix 
$$\op_{N}^{(k)}=\tr_{k+1\dots N}\(\op_N\)$$ can be expressed in terms of $\op_{N}^{(1)}$ as 
\begin{equation*}
\op_{N}^{(k)}(x_1,\dots,x_k,y_1,\dots,y_k)=\sum_{\pi\in S_k} {\rm sgn}(\pi)\prod_{j=1}^k \op_{N}^{(1)}(x_j,y_{\pi(j)}),
\end{equation*}
where $S_k$ is the set of permutations of $k$ elements and ${\rm sgn}(\pi)$ is the sign of the permutation $\pi$.
We distinguish two cases:
\begin{itemize}
\item[i)] if $\boldsymbol{\rho}_N$ has rank one, i.e.~$\op_N=|\psi_N\rangle\langle\psi_N|$ is a rank one projection, we say that $\boldsymbol{\rho}_N$ represents a pure state;
\item[ii)] if $\boldsymbol{\rho}_N$ has rank strictly bigger than one, we say that $\boldsymbol{\rho}_N$ represents a mixed state.
\end{itemize}

\noindent In both cases, the time evolution of the quasi-free state $\op_N$ is denoted by $\boldsymbol{\rho}_{N,t}$ and it solves the Liouville--von Neumann equation
\begin{equation}\label{eq:LvN}
i\,\hbar\,\partial_t\,\boldsymbol{\rho}_{N,t}=\left[H_N\,,\,\boldsymbol{\rho}_{N,t}\right]
\end{equation}
with initial datum $\boldsymbol{\rho}_N$.
Here $\left[H_N,\boldsymbol{\rho}_{N,t}\right]$ denotes the commutator between the self-adjoint operator
\[
H_{N}=\sum_{j=1}^N-\frac{\,\hbar^2}{2}\Delta_{x_j}+\frac{1}{N}\sum_{1\leq j<k\leq N}K(x_j-x_k)
\] 
and $\boldsymbol{\rho}_{N,t}$. \\
To better understand the structure of pure and mixed states, we introduce the Wigner transform, that is a transformation associating to a the kernel of a one-particle density matrix a function on the one-particle phase space. For a one-particle operator $\op$, we denote its Wigner transform by $f_{\op}$ and
\[\label{eq:wigner}
f_{\op}(x,v)=\int e^{-y\cdot v/\hbar}\op\(x+\frac{y}{2},x-\frac{y}{2}\)\,dy.
\]
Observe moreover that $f_{\op}$ is normalized to 1, i.e.
\begin{equation*}
\int_{\R^6}f_{\op}(x,v)\,dx\,dv=\hbar^3\tr(\op)=1.
\end{equation*}
If $\op$ is a pure state, hence an orthogonal projection onto the subspace of $L^2(\mathbb{R}^3)$ spanned by $\{f_j\}_{j=1}^N$, then its Wigner transform $f_{\op}$ converges, as $\hbar$ goes to zero, to the characteristic function of a set, and hence enjoys the same regularity properties of the characteristic functions. In particular, $f_{\op}$ can be an element in $W^{1,1}(\mathbb{R}^6)$, but not in $W^{s,p}(\mathbb{R}^6)$ with $s,p>1$ (see \cite{lafleche_optimal-regularity_2023}). 

\noindent In the case of mixed states, we can assume more regularity, such has $W^{s,p}$ Sobolev regularity with $s,p>1$, whenever the kernel of the operator $\op$ has a smooth kernel. \\
Thus from the one hand, the friendly structure $\op^2=\op$ can be used when dealing with pure states, whereas for mixed states $\op$ is only known to be a non-negative bounded operator; on the other hand we cannot rely on regularity properties of $f_{\op}$ if $\op$ is a pure state.

\subsection{Effective equations}

For $N$ sufficiently large, we expect $\op_{N,t}^{(1)}$, the time evolution of $\op_{N}^{(1)}$, to converge in some  topology to a solution $\op$ to the time-dependent Hartree-Fock equation
\begin{equation}\label{eq:HF}
i\hbar\partial_t\op=[-\hbar^2\Delta+V_{\op}-X_{\op},\op]
\end{equation}
where $\op$ is a time-dependent, nonnegative, self-adjoint and trace class operator acting on $L^2(\mathbb{R}^3)$, satisfying 
\begin{equation}\label{eq:normalization}
\tr(\op)=\hbar^{-3}
\end{equation}
 and $0\leq\op\leq 1$, $V_{\op}$ is the multiplication operator by $K*\rho_t$, representing the mean-field potential with spatial density $$\rho(t,x)=\diag(\op)(x):=\hbar^3\op(x,x),$$ and $X_{\op}$ is the exchange term defined in terms of its integral kernel 
$X_{\op}(x,y)=K(x-y)\op(x,y)$. As we did for the reduced density matrices, with an abuse of notation we denote here  by $\op(x,y)$ the kernel of the operator $\op$.\\
Notice that the choice of the normalization \eqref{eq:normalization} ensures that $\int_{\R^3} \rho_t(x)\,dx=\hbar^3\tr(\op)=1$, that makes the comparison with probability densities on the phase space more transparent.
In particular observe that equation \eqref{eq:HF} still depends on the number of particles $N$ through $\hbar$ and $\op$, thus claiming for an investigation of the limit $\hbar\to 0$, or $N\to\infty$. In fact, considering the Wigner transform $f_{\op_t}$ of the operator $\op_t$ solution to \eqref{eq:HF} yields
\begin{equation}\label{eq:heuristic}
\begin{split}
i\hbar\partial_t\,f_{\op}&=\int i\hbar\partial_t\op\(x+\frac{y}{2},x-\frac{y}{2}\)\,e^{-iv\cdot y/\hbar}dy\\
&=\int \frac{1}{2}\(-\Delta_{x+\frac{y}{2}}+\Delta_{x-\frac{y}{2}}\)\,\op\(x+\frac{y}{2},x-\frac{y}{2}\)\,e^{-iv\cdot y/\hbar}\,dy\\
&\ +\int \left[V_{\op}\(x+\frac{y}{2}\)-V_{\op}\(x-\frac{y}{2}\)\right]\,\op\(x+\frac{y}{2},x-\frac{y}{2}\)\,e^{iv\cdot y/\hbar}\,dy.
\end{split}
\end{equation}
We observe that $-\Delta_{x+\frac{\hbar y}{2}}+\Delta_{x-\frac{\hbar y}{2}}=-2\hbar\nabla_x\cdot\nabla_y$ and $$\left[V_{\op}\(x+\frac{\hbar y}{2}\)-V_{\op}\(x-\frac{\hbar y}{2}\)\right]\simeq \hbar\,y\cdot\nabla V_{\op}+O(\hbar^2).$$ Changing variable $y\to \hbar y$,  dividing by $i\hbar$ and integrating in $y$, we get
\begin{equation}\label{eq:approx-vlasov}
\partial_t\,f_{\op}=-\nabla_x f_{\op_t}+\nabla V_{\op}\cdot\nabla_v f_{\op} +O(\hbar).
\end{equation}
This heuristic computation suggests that, in the limit of large $N=\hbar^{-3}$, $f_{\op}$ approaches a solution to the Vlasov equation
\begin{equation}\label{eq:Vlasov}
\partial_t f+v\cdot\nabla_x f-\nabla(K*\rho_f)\cdot\nabla_v f=0,
\end{equation} 
where $f=f(t,x,v)$ is the classical phase space distribution of particles, $\nabla K*\rho_f$ is the self-induced force field and $\rho_f$ is the spatial density defined as $\rho_f(t,x)=\int f(t,x,v)\,dv$.  Equation \eqref{eq:Vlasov} describes the classical dynamics of a large number of interacting non collisional particles subject to many weak interactions whose collective effect can be approximated by an averaged mean-field potential. The Vlasov equation is much used in plasma physics and astrophysics as its description is particularly suited for plasmas and dense gases.  Observe that $\op$ and $f$ are time-dependent quantities. When dealing with the initial data we will use the notation $\op^{\rm in}=\op|_{t=0}$ and $f^{\rm in}=f|_{t=0}$.

\subsection{State of the art}

{The pioneering works \cite{Narnhofer-Sewell_1981} and \cite{Spohn_1981} provide a
 first rigorous derivation of the Vlasov equation~\eqref{eq:Vlasov} from the $N$-body Schr\"{o}dinger equation~\eqref{eq:MBS} in a combined mean-field and semiclassical limit in the case of analytic and twice differentiable potentials respectively. This approach has been later reconsidered in~\cite{Graffi-Martinez-Pulvirenti_2003,Chen-Lee-Liew_2021}. The approximation of the many-body Schr\"{o}dinger equation in terms of the Hartree and Hartree-Fock equations, when  $N=\hbar^{-3}$ is large but finite, has been considered in \cite{Elgart-Erdos-Schlein-Yau_2004} for analytic potentials and for short times, and later extended to arbitrarily large fixed times and to smooth potentials in \cite{Benedikter-Porta-Schlein_2014}, where explicit rates of convergence were obtained by means of a new method based on second quantization techniques reminiscent of \cite{Ginibre-Velo_1985,Rodianski-Schlein_2009} and the semiclassical structure of the initial data \begin{equation}\label{eq:semicl-regular}
 \hbar^{3}{\rm Tr}\,\left|\com{\frac{x}{i\hbar},\op}\right|\leq C,\quad\quad  \hbar^{3}{\rm Tr}\,|[\nabla,\op]|\leq C
\end{equation}
is exploited. 
When referring to \eqref{eq:semicl-regular}, the name semiclassical is due to the fact that the commutator between $\op$ and the position operator (or the momentum operator respectively) is small as $\hbar$ goes to zero, and therefore they almost commute. }{In the same spirit of \cite{Benedikter-Porta-Schlein_2014}, partial results have been obtained for singular interactions in \cite{Porta-Rademacher-Saffirio-Schlein_2017,Saffirio_2018}, where the convergence of the many-body fermionic dynamics to the Hartree equation with inverse power law potential (including Coulomb) has been proven for translation invariant states, close to a Slater determinant.\\  
The same problem has been studied in different regimes in \cite{Erdos-Yau_2001,Bardos-Golse-Gottlieb-Mauser_2003, Frohlich-Knowles_2011, Bach-Breteaux-Petrat-Pickl-Tzaneteas_2016,  Petrat-Pickl_2016} for the Coulomb interaction. Recently, new techniques reminiscent of the ones used in the mean-field limit for systems of classical particles have been developed in \cite{Golse-Mouhot-Paul_2016,Golse-Paul_2017,Golse-Paul-Pulvirenti_2018,Golse-Paul_2019}. Once the validity of the Hartree-Fock approximation is established, one can investigate its classical limit $\hbar=N^{-1/3}\to 0$. In the semiclassical regime, the convergence of the Hartree dynamics towards the Vlasov equation was proven in \cite{Lions-Paul_1993} in weak topology, including singular potentials such as the Coulomb interaction, using compactness methods. Explicit rates in stronger topologies were then obtained in \cite{Athanassoulis-Paul-Pezzotti-Pulvirenti_2011, Amour-Khodja-Nourrigat_2013,Benedikter-Porta-Saffirio-Schlein_2016} for regular potentials, in \cite{Saffirio_2019,Saffirio_2020} for a certain class of singular potentials and in \cite{Golse-Paul_2017,Lafleche_2019, Lafleche_2021} for regular and singular interactions in weak topology. Notice also that the study of the classical limit of infinite gases has been addressed in \cite{Lewin-Sabin_2020}  for local perturbations of stationary states.\\}
Although many progresses have been done in recent years, the most relevant (from a physics viewpoint) cases in which particles interact through the Coulomb or gravitational potentials are still out of reach.  An attempt was done in
  \cite{Lafleche-Saffirio_2020,Chong-Lafleche-Saffirio_2021,Chong-Lafleche-Saffirio_2022}, where the $3d$ Vlasov equation has been derived from a system of $N$ fermions interacting through an inverse power law potential of the form 
\begin{equation}\label{eq:interaction}
K(x)=\pm\frac{1}{|x|^a}, \quad x\in\mathbb{R}^3.
\end{equation}
The derivation holds for any bounded but arbitrarily large time interval if $a\in[0,\frac{1}{2})$, and for times of order $\sqrt{\hbar}$ if $a\in[\frac{1}{2},1]$.\\

\subsection{Notations and main result}

In these notes we will review the results obtained in \cite{Lafleche-Saffirio_2020} and \cite{Chong-Lafleche-Saffirio_2021}.
To this end, we introduce some notations.
For a one-particle self-adjoint operator $\op$ acting on $L^2(\mathbb{R}^3)$, let $\L^p$ be the semiclassical analogue of Lebesgue spaces equipped with the rescaled Schatten norm 
\begin{equation*}
\Nrm{\op}{\L^p}=\hbar^\frac{3}{p}\Nrm{\op}{p}=\hbar^\frac{3}{p}\tr(|\op|^p)^\frac{1}{p},
\end{equation*}
and let $\L^\infty$ be the space of bounded operators equipped with the operator norm $\Nrm{\op}{\L^\infty}$. For ${\rm m}:=1+|\opp|^n$, where $\opp=-i\hbar\nabla$ is the momentum operator, we define the weighted semiclassical Lebesgue norms by $\Nrm{\op}{\L^p({\rm m})}:=\Nrm{\op\,{\rm m}}{\L^p}$ and the semiclassical analogue of Sobolev norms by
\begin{equation*}
\displaystyle
\begin{array}{ll}
\Nrm{\op}{\mathcal{W}^{1,p}({\rm m})}^p:= \Nrm{\op}{\L^p({\rm m})}^{p}+\sum\limits_{j=1}^3 \Nrm{\Dhvj\op}{\L^p({\rm m})}^p+\Nrm{\Dhxj\op}{\L^p({\rm m})}^p & p\in[1,\infty),\\\\
\Nrm{\op}{\mathcal{W}^{1,\infty}({\rm m})}:= \Nrm{\op}{\L^\infty({\rm m})}+\sup\limits_{j=1,2,3}\(\Nrm{\Dhvj\op}{\L^\infty({\rm m})}+\Nrm{\Dhxj\op}{\L^\infty({\rm m})}\) & p=\infty,
\end{array}
\end{equation*}
where 
\begin{equation}\label{eq:quantum-gradients}
\Dhx\op:=\com{\nabla,\op}\quad\mbox{ and }\quad\Dhv\op:=\com{\frac{x}{i\hbar},\op}.
\end{equation}
For any integrable function $f$ on the phase space, we introduce the Weyl quantization $\op_f$, defined as the operator with integral kernel
\begin{equation*}
\op_f(x,y)=\int_{\R^3} e^{-2i\pi(y-x)\cdot v}f\(\frac{x+y}{2},\hbar\,v\)\,dv.
\end{equation*}
To describe the many-body fermionic system, we introduce the Hilbert space $\mathfrak{h}:=L^2(\R^3)$ and the $n$-fold antisymmetric tensor product of $\mathfrak{h}$, $\mathfrak{h}^{\wedge n}=\mathfrak{h}\wedge\dots\wedge\mathfrak{h}$. We define the fermionic Fock space over $\mathfrak{h}$ by 
\begin{equation*}
\mathcal{F}:=\mathbb{C}\oplus\bigoplus_{n=1}^\infty \mathfrak{h}^{\wedge n}
\end{equation*}
equipped with the norm induced by the inner product on $\mathcal{F}$.\\ Let
\begin{equation*}
\mathcal{N}\psi=\(n\psi^{(n)}\)_{n\in\N}
\end{equation*}
be the number of particles operator on $\mathcal{F}$, where $\psi^{(n)}$ is the $n$-particle sector of the Fock space vector $\psi\in\mathcal{F}$. Let $\L^p(\mathcal{F})$ be the semiclassical Lebesgue spaces on the Fock space with norm $\Nrm{\op_N}{\L^p(\mathcal{F})}:=\hbar^\frac{3}{p}\tr(|\op_N|^p)^\frac{1}{p}$, so that $\Nrm{\op_N}{\L^1(\mathcal{F})}=1$ and $\Nrm{\mathcal{N}\op_N}{\L^1(\mathcal{F})}=N$.\\
Hereafter we will consider a situation in which the following assumptions are satisfied:
\begin{itemize}
\item[(A1)] {\it Normalization constraints.} Let $\op$ be a bounded operator satisfying 
\begin{equation*}
\Nrm{\op}{\L^\infty}=C_{\infty},\quad\quad \tr(\op)=\hbar^{-3}
\end{equation*}
for some constant $C_{\infty}>0$.
\item[(A2)] {\it Regularity of $\op$, uniform in $\hbar$.} Let $\op$ be a nonnegative operator satisfying 
\begin{equation*}
\begin{split}
&\op\in\mathcal{W}^{2,2}({\rm m})\cap\mathcal{W}^{2,4}({\rm m})\\
&\sqrt{\op}\in\mathcal{W}^{1,2}({\rm m})\cap\mathcal{W}^{1,q}({\rm m})
\end{split}
\end{equation*}
with $q\in\left[\frac{6}{1-2a},\infty\right]$, where $a$ is the power of the singular interaction. 
\item[(A3)] {\it Propagation of moments and regularity of $f$.} Let $f$ be a function on the phase space such that
\begin{equation*}
(1+|x|^{8}+|v|^{8})\nabla_x^{\ell}\nabla_v^{m}f \in L^\infty(\R^6)\cap L^2(\R^6),\quad\quad\ell+m\leq 9.
\end{equation*}
\end{itemize}

We are now ready to state our result.

\begin{theorem}\label{thm:combined-MF-semiclassical}
Let $a\in(0,\frac{1}{2})$ in \eqref{eq:interaction}, $n\in 2\N$ such that $n>2$ and $\op$ be a solution to the Hartree-Fock equation \eqref{eq:HF} with initial datum $\op^\init\in\L^\infty({\rm m})$ satisfying assumptions {\rm (A1)} and {\rm (A2)}. Let $f$ be a nonnegative solution to the Vlasov equation \eqref{eq:Vlasov} with initial datum $f^\init$ satisfying {\rm (A3)}.   Let $\op_N$ be a solution to the Liouville-von Neumann equation \eqref{eq:LvN} with initial condition $\op_N^\init$ such that $\op_N^\init\in\L^1(\mathcal{F})$ and $\com{\mathcal{N},\op_N^\init}=0$.\\ Then for every $T>0$ there exist an operator $\op_{N,f}^\init\in\L^1(\mathcal{F})$ and a constant $C_T>0$ such that
\begin{equation}\label{eq:main-estimate}
\Nrm{\op_{N}^{(1)}-\op_f}{\L^1}\leq C_T\(\frac{1}{N}+\hbar\)\(1+\Nrm{(\mathcal{N}+N)^k(\op_N^\init+\op_{N,f}^\init)}{\L^1(\mathcal{F})}\),
\end{equation} 
for $k\geq \frac{1}{2}+\frac{3}{2}\lceil\frac{\ln N}{\ln(N\,\hbar^2)}\rceil$.
\end{theorem} 
\begin{remark}
\item[i)] Equation~\eqref{eq:scaling} yields $\hbar$ as leading order in the approximation for $N$ large. This matches the rate obtained in the heuristic computation \eqref{eq:heuristic}.
\item[ii)] Recalling the definition of quantum gradients \eqref{eq:quantum-gradients}, we can read (A2) as a generalisation of \eqref{eq:semicl-regular}. Thus for singular interactions more ``quantum integrability'' is needed on the initial states with respect to the case of smooth potentials. 
\item[iii)] Consider the case $a=1$, i.e. the Coulomb potential, and let
 \begin{equation*}
 K_R(x)=\int_0^{R^{-2}}\frac{e^{-\pi|x|^2}s}{\sqrt{s}}ds\longrightarrow\frac{1}{|x|}\quad \mbox{ as } R\to 0
 \end{equation*}
 be a cut-off Coulomb potential. Then
 \begin{equation*}
 \Nrm{\op_{N}^{(1)}-\op_f}{\L^1}\leq \frac{C_T\,e^{t/\sqrt{R}}}{\sqrt{N}} + C_T\,\hbar,
 \end{equation*}
with $k\geq \frac{1}{2}+\frac{3}{2}\lceil\frac{\ln N}{\ln(N\,\hbar^2)}\rceil$. Thus the convergence still holds true on times $t\ll \sqrt{\hbar}$. This is an improvement with respect to previous results (see e.g.~\cite{Petrat-Pickl_2016}), where the convergence was obtained on a time scale $t\ll \hbar$. 
\item[iv)] Furthermore, using Theorem~\ref{thm:combined-MF-semiclassical} and \cite[Theorem~1.1]{chong_optimal_2022}, we get an analogue of Theorem~\ref{thm:combined-MF-semiclassical} in Hilbert-Schmidt norm:
\begin{equation*}
\Nrm{\op_{N}^{(1)}-\op_f}{\L^2}\leq C_T\(\frac{1}{\sqrt{N}}+\hbar\)\(1+\Nrm{(\mathcal{N}+N)^k(\op_N^\init+\op_{N,f}^\init)}{\L^1}\).
\end{equation*}
Notice that if $\(1+\Nrm{(\mathcal{N}+N)^k(\op_N^\init+\op_{N,f}^\init)}{\L^1}\)\leq C$, we obtain convergence in $L^2$ for the functions on the phase space using that $\Nrm{\op_f}{\L^2}=\Nrm{f}{L^2}$, thus proving the quantitative bound
\begin{equation*}
\Nrm{f_{N}^{(1)}-f}{L^2(\R^6)}\leq C_T\(\frac{1}{\sqrt{N}}+\hbar\),
\end{equation*}
where $f_{N}^{(1)}:=f_{\op_N^{(1)}}$ is the Wigner transform of the one-particle reduced density matrix $\op_{N}^{(1)}$.
\end{remark}


\section{Derivation of the Hartree Equation}\label{sec:MF}

{\bf Step 1. Purification.}
The very first difficulty we encounter arises from considering mixed states instead of pure states. For a spectral set $\{\lambda_j,\psi_j\}_{j\geq 0}$, $\lambda_j\in[0,1]$ and $\psi_j\in\mathcal{F}(\mathfrak{h})$, we can express $\op_N$ as
\begin{equation*}
\op_N=\sum_{j\geq 0}\lambda_j|\psi_j\rangle\langle\psi_j|,
\end{equation*}
that in general it is not a rank one projection. However, we can see it as a pure state on a larger Fock space by observing that
\begin{equation*}
\op_N^{\frac{1}{2}}=\sum_{j\geq 0} \lambda_j^{\frac{1}{2}}|\psi_j\rangle\langle\psi_j|\simeq \sum_{j\geq 0}\lambda_j^\frac{1}{2}\psi_j\otimes\overline{\psi_j}\in\mathcal{F}\otimes\mathcal{F}
\end{equation*}
and by noticing that there exists $U$ isomorphism such that 
\begin{equation}\label{eq:iso}
\mathcal{F}(\mathfrak{h})\otimes\mathcal{F}(\mathfrak{h})\simeq_U \mathcal{F}(\mathfrak{h}\oplus\mathfrak{h})=:\mathcal{G}.
\end{equation}
This simple observation is the key to recast the problem for mixed states to a Cauchy problem for states that exhibit the structure of pure states in the Fock space on the larger double Hilbert space $\mathcal{G}$ (see \cite{Araki,Derezinski,Benedikter-Jaksic-Porta-Saffirio-Schlein_2016}). \\
On $\mathcal{G}$ we introduce the left and right creation and annihilation operators as follows: for every $f\in\mathfrak{h}$, the left and right creation  operators are
\begin{equation*}
\begin{array}{ll}
a_l^*(f):=a^*(f\oplus 0), & a_r^*(f):=a^*(0\oplus f),
\end{array}
\end{equation*}
and the left and right annihilation operators are
\begin{equation*}
\begin{array}{ll}
a_l(f):=a(f\oplus 0), & a_r(f):=a(0\oplus f),
\end{array}
\end{equation*}
where $a$ and $a^*$ are the usual annihilation and creation operators on $\mathcal{F}(\mathfrak{h})$, satisfying the canonical anticommutation relations.  Moreover, for an observable $J$ with distributional kernel $J(x,y)$, we define the left and right second quantization of $J$ by
\begin{equation*}
\begin{array}{l}
d\Gamma_l(J):=d\Gamma(J\oplus 0)=\int_{\R^6}J(x,y)\,a_{x,l}^*\,a_{y,l}\,dx\,dy\\
d\Gamma_r(J):=d\Gamma(0\oplus J)=\int_{\R^6}J(x,y)\,a_{x,r}^*\,a_{y,r}\,dx\,dy
\end{array}
\end{equation*}
respectively, where $a_{z,l}$ and $a_{z,r}$ are the left and right annihilation operators at the position $z$, and $a_{z,l}^*$ and $a_{z,r}^*$ are their adjoints and we will refer to them as the left and right creation operator-valued distributions at the position $z$, respectively. The number of particles operator is then defined as the quantization of the identity, i.e.
\begin{equation*}
\mathcal{N}:=\mathcal{N}_l+\mathcal{N}_r=d\Gamma_l(1)+d\Gamma_r(1),
\end{equation*} 
where $d\Gamma_l(1)=d\Gamma(1\oplus 0)$ and $d\Gamma_r(1)=d\Gamma(0\oplus 1)$.\\
These notations allows us to rewrite the solution of \eqref{eq:LvN} with initial datum $\op_N^\init$  in the interaction picture
\begin{equation*}
\op_N=e^{-iH_N t/\hbar}\op_N^\init e^{iH_N t/\hbar}
\end{equation*}
as a vector $\Phi(t)\in\mathcal{G}$ as follows:
\begin{equation*}
\Phi(t):=e^{-iL_N t/\hbar}\Phi^\init
\end{equation*}
where $$L_N=U(H_N\otimes 1-1\otimes H_N)U^*,$$ with $U$ the isomorphism given in \eqref{eq:iso}. Hence, we can write the one-particle reduced density matrix of $\op_N$ in terms of $\Phi(t)$ as
\begin{equation}\label{eq:1pdm}
\op_{N:1}(x,y)=\langle\Phi(t),\,a^*_{x,l}\,a_{y,l}\,\Phi(t)\rangle.
\end{equation}
\smallskip

{\bf Step 2. Bogoliubov transformation.}
The reason to adopting the second quantization formalism for this problem is two-fold. On the one hand we want to quantify the difference between the dynamics in terms of the fluctuations around the limiting equation (see step 3 below). To this end, working in second quantization with no fixed number of particles helps. We can indeed think of the Fock space as the quantum analogue of the grand canonical ensamble in classical statistical mechanics. On the other hand on the Fock space $\mathcal{G}$ we can define the so-called Bogoliubov transformation, allowing for a representation of a quasi-free mixed state as a rotation of the vacuum in the Fock space $\mathcal{G}$. 
The advantage of this tool is that this transformation acts as a time-dependent change of variables that transforms the reference frame. Choosing the reference frame to be the one of $\op$, solution to \eqref{eq:HF}, allows us to cancel several terms. Indeed, let
\begin{equation*}
u=\sqrt{1-\op}\quad \mbox{ and }\quad v=\sqrt{\op},
\end{equation*}
and construct a unitary map $R_{\op}:\mathcal{G}\to\mathcal{G}$ such that 
\begin{equation}\label{eq:bog-properties}
R^*_{\op}\,a_{x,l}\,R_{\op}=a_l(u_x)-a_r^*(\overline{v}_x),\quad
R^*_{\op}\,a_{x,r}\,R_{\op}=a_r(\overline{u}_x)+a_l^*({v}_x),
\end{equation}
where we used the notation
\begin{equation*}
u_x(y)=u(y,x),\quad\quad v_x(y)=v(y,x).
\end{equation*}
Notice that $u,\,v$ are well defined because $\op$ is a fermionic operator, i.e.~$0\leq \op\leq 1$. Moreover, since $\tr(\op)=\hbar^{-3}$, the $p$-Schatten norms of $\op$ and $v$ are finite, for $p\in[1,\infty]$. However $u$, despite being bounded in $\L^\infty$, is not bounded in other Schatten norms and this makes the analysis more delicate.\\
The choice of $R_{\op}$ allows us to construct a quasi-free state with one-particle reduced density matrix $\op$ on $\mathcal{G}$. To this end, let $\Phi_{\op}$ be the rotation of the vacuum $\Omega_{\mathcal{G}}\in\mathcal{G}$ by the Bogoliubov transformation: 
\begin{equation}\label{eq:new-vacuum}
\Phi_{\op}=R_{\op}\Omega_{\mathcal{G}}\in\mathcal{G}.
\end{equation}
This construction is known as Araki-Wyss representation in quantum statistical mechanics {(see for instance \cite{Araki,Derezinski})}.\\
Then, it is readily seen that the one particle reduced density matrix associated with $\Phi_{\op}$ is
\begin{equation}\label{eq:structure}
\begin{split}
\langle\Phi_{\op},\,a^*_{l,y}\,a_{l,x}\,\Phi_{\op} \rangle&=\langle\Omega,\,R_{\op}\,a^*_{l,y}\,R_{\op}\,R_{\op}^*\,a_{l,x}\,R_{\op}\,\Omega \rangle \\
&=\langle\Omega,\,(a_l^*(u_y)-a_r(\overline{v_y}))\,(a_l(u_x)-a_r^*(\overline{v_x})\,\Omega\rangle\\
&=\langle\Omega,\,a_r(\overline{v_y})\,a_r^*(\overline{v_x})\,\Omega\rangle\\
&=(v^*\,v)(x,y) \\
&=\op(x,y),
\end{split}
\end{equation}
where we used the relations \eqref{eq:bog-properties}.
\smallskip

{\bf Step 3. Fluctuation dynamics.}
We are interested in the time evolution of $R_{\op}$. Clearly for positive times equation \eqref{eq:structure} does not hold because the interaction among the fermions creates correlations. However we expect that for weakly interacting fermions equation \eqref{eq:structure} is approximately true. Indeed, using \eqref{eq:new-vacuum}, for $\Psi_{\rm fluct}\in\mathcal{G}$ defined as
\begin{equation*}
\Psi_{\rm fluct}:=R^*_{\op}\Phi(t)=R^*_{\op}e^{-iL_N t/\hbar}R_{\op^\init}\Omega,
\end{equation*}
and $\op_{N:1}$ as in Theorem~\ref{thm:combined-MF-semiclassical}, we can estimate the error in the mean-field approximation by the mean number of particles of the fluctuation dynamics around a quasi-free state, i.e.
\begin{equation}\label{eq:densities-number-particles}
\Nrm{\op_{N:1}-\op}{\L^1}\leq\frac{C}{\sqrt{N}}\Nrm{(\mathcal{N}+1)^{\frac{1}{2}}\Psi_{\rm fluct}}{\mathcal{G}}^2.
\end{equation} 
To obtain \eqref{eq:densities-number-particles}, we notice that
\begin{equation*}
\begin{split}
\op_{N:1}(x,y)
&=\op(x,y)+\langle \Phi(t),\,a^*_{y,l}\,a_{x,l}\,\Phi(t) \rangle\\
&=\langle\Psi_{\rm fluct},\,R^*_{\op}\,a^*_{y,l}\,R_{\op}\,R^*_{\op}\,a_{x,l}\,R_{\op}\,\Psi_{\rm fluct}\rangle.
\end{split}
\end{equation*}
By \eqref{eq:bog-properties} and for any observable $J$, we get
\begin{equation*}
\begin{split}
\tr(J&(\op_{N:1}-\op))\\
&=\langle \Psi_{\rm fluct},\,\(d\Gamma_l(u\,J\,u)-d\Gamma_r(\overline{v}\,J\,v)-d\Gamma^+_{l,r}(v\,J\,u)-d\Gamma^-_{r,l}(v\,J\,u)\)\,\Psi_{\rm fluct} \rangle
\end{split}
\end{equation*}
where $$d\Gamma^{+}_{\sigma,\sigma'}(J)=\int J(x,y)a_{x,\sigma}^*a_{y,\sigma'}^*dxdy,\mbox{ and } d\Gamma^{-}_{\sigma,\sigma'}(J)=\int J(x,y)a_{x,\sigma}a_{y,\sigma'}dxdy.$$ 
Using that $\|u\|_{\L^\infty}\leq 1$ and $\|v\|_{\L^\infty}\leq 1$, and that $d\Gamma(J)$ can be bounded in terms of the number of particles operator, we get
\begin{equation*}
\tr(J(\op_{N:1}-\op))\leq C\Nrm{J}{\L^\infty}\Nrm{(\mathcal{N}+1)^\frac{1}{2}\Psi_{\rm fluct}}{\mathcal{G}}
\end{equation*}
that by duality yields \eqref{eq:densities-number-particles}. Whence, to control the mean-field approximation error, we need to bound the expectation of the number of particles in the fluctuation state $\Psi_{\rm fluct}$. 
\smallskip

{\bf Step 4. Control on the growth of $\mathcal{N}$ in the fluctuation state.} Let
\begin{equation*}
\mathcal{U}(t,s):=R_{\op}^*\,e^{-iL_N t/\hbar}\,R_{\op^\init}
\end{equation*}  
be the unitary two-parameter semigroup with generator $G_t$ satisfying
\begin{equation*}
i\hbar\partial_s\,\mathcal{U}(t,s)=G_t\,\mathcal{U}(t,s),\quad\quad \mathcal{U}(s,s)=1,
\end{equation*} 
where the generator $G_t$ is given by
\begin{equation*}
G_t=d\Gamma_l(H_{\op})-d\Gamma_r(\overline{H}_{\op})+D+(Q+\widetilde{Q}+{\rm h.c.}),
\end{equation*}
and
\begin{equation}\label{eq:HF-hamiltonian}
H_{\op}=-\hbar^2\Delta+V_{\op}-X_{\op}
\end{equation}
 is the Hartree-Fock Hamiltonian, $D$ contains terms that commute with $\mathcal{N}$, $Q$ and $\widetilde{Q}$ contain terms that do not commute with the number operator. We refer to $\mathcal{U}(t,s)$ as the fluctuation dynamics and observe that, by definition of $\mathcal{U}(t,s)$, $\Psi_{\rm fluct}=\mathcal{U}(t,0)\,\Omega$. \\
 In step 3 we highlighted that to give an explicit bound on the mean-field approximation error we need to bound the number of particles in the state $\Psi_{\rm fluct}$. In other words, we consider 
\begin{equation}\label{eq:gronwall}
i\hbar\partial_t\,\langle\mathcal{U}(t,0)\psi,\,(\mathcal{N}+N)\,\mathcal{U}(t,0)\,\psi\rangle
\end{equation}
for $\psi\in\mathcal{G}$ and bound it by means of Gr\"{o}nwall's Lemma. The result is the following 
\begin{proposition}\label{prop:bound-calN}
For $k_0,k>0$ and $\psi\in\mathcal{G}$, it holds
\begin{equation*}
\Nrm{(\mathcal{N}+N)^{k_0}\mathcal{U}(t,0)\,\psi}{\mathcal{G}}\leq C\,e^{\lambda\,t}\(\Nrm{(\mathcal{N}+N)^{k_0+\frac{3}{2}k}\psi}{\mathcal{G}}+\frac{\hbar^\frac{k}{2}t}{N^{\frac{k}{2}-k_0}}\Nrm{(\mathcal{N}+N)^{\frac{3}{2}k}\psi}{\mathcal{G}}\),
\end{equation*}
where $\lambda$ depends on $\Nrm{\Dhv\sqrt{\op}\,{\rm m}}{\L^{q_0}}$ and $\Nrm{\Dhv\sqrt{\op}\,{\rm m}}{\L^{q_1}}$, $1\leq q_0<q_1<\infty$ such that $\frac{1}{2}\(\frac{1}{q_0}+\frac{1}{q_1}\)=1-\frac{3}{a+1}$. 
\end{proposition}

\noindent Notice that $\Nrm{\Dhv\sqrt{\op}\,{\rm m}}{\L^{q}}$ can be bounded uniformly in $\hbar$ (see \cite[Part II]{Chong-Lafleche-Saffirio_2021} and \cite{Chong-Lafleche-Saffirio_2022}).\\
To prove Proposition~\ref{prop:bound-calN} we show that each term in the generator $G_t$ that do not commute with $\mathcal{N}$ is bounded uniformly in $\hbar$. However, two difficulties arise:

${\rm (i)}$ when $i\hbar\partial_t$ acts on $\mathcal{U}(t,s)$ in \eqref{eq:gronwall}, we are lead to bound the commutator between the number of particles operator $\mathcal{N}$and the generator $G_t$, which contains terms where the singular interaction appears;

${\rm (ii)}$ we need to cancel the $\hbar$ on the right-hand side of \eqref{eq:gronwall}, and therefore to exploit the hidden commutator structure, in particular the fact that $[u,v]=0$. \\
Using the fact that some terms in the generator $G_t$ commute with $\mathcal{N}$, we are left with the control of the terms in the generator that do not commute with the number of particle operator, namely $\widetilde{Q}$ and $Q$ (quartic terms in the creation and annihilation operators). We first focus on 
\begin{equation*}
\widetilde{Q}=\frac{1}{2N}\int_{\R^3}d\Gamma_{l,r}^+(uK_xv)\,d\Gamma_{l,r}^+(u\delta_xv)-d\Gamma_{l,r}^+(v\delta_xu)\,d\Gamma_{l,r}^+(uK_xv)\,dx,
\end{equation*}
where $u\delta_xv$ denotes the operator with kernel $(u\delta_xv)(y,z)=u(y,x)v(x,z)$ and $K_x(y)=K(x-y)$.
The term in which $\widetilde{Q}$ appears can be bounded uniformly in $\hbar$ by exploiting the commutator structure
\begin{equation*}
uK_xv=vK_xu+u\com{K_x,v}-v\com{K_x,u}.
\end{equation*}
The contributions of $Q$ are more difficult to handle. They are of the form
\begin{equation*}
\frac{1}{N}\iint_{\R^6}K(x-y)\,a_{l}^*(u_x)\,a_{r}^*(\overline{v}_x)\,a_{l}^*(u_y)\,a_{l}(u_y)\,dx\,dy.
\end{equation*}
In order to exploit the hidden commutator structure we further decompose this term using that $\com{u,v}=0$. Combining all the terms we obtain the decomposition
\begin{equation*}
Q=(P+\widetilde{P}+{\rm h.c.}).
\end{equation*}
The terms in $\widetilde{P}$ do not present cancellations and we will deal with them in Step 5 below. The terms in $P$ instead are the ones responsible for the restriction to inverse power law potentials with $a<\frac{1}{2}$.
They are of the form
\begin{equation*}
\mathfrak{P}=\frac{1}{N}\int_{\R^3}a_{l}^*(u_x)\,a_{r}^*(\overline{v}_x)\,d\Gamma_l(K_x)\,dx.
\end{equation*}
By the Cauchy-Schwarz inequality, for every $\psi_1,\,\psi_2\in\mathcal{G}$ we get
\begin{equation*}
\begin{split}
|\langle\psi_1,\mathfrak{P}\psi_2\rangle|&\leq \frac{1}{N}\(\int_{\R^3}\Nrm{a_{l}(u_x)\,\psi_1}{\mathcal{G}}^2\)^{\frac{1}{2}}
\(\int_{\R^3}\Nrm{a_{r}^*(\overline{v}_x)\,d\Gamma_l(K_x)\,\psi_2\,}{\mathcal{G}}^2\)^{\frac{1}{2}}\\
&\leq \frac{1}{N}\Nrm{\mathcal{N}_l^{\frac{1}{2}}\psi_1}{\mathcal{G}}\(\int_{\R^3}\rho(x)\,\Nrm{d\Gamma_l(K_x)\,\psi_2}{\mathcal{G}}^2dx\)^\frac{1}{2}.
\end{split}
\end{equation*}
where in the last inequality we used that $\Nrm{v_x}{L^2}=N\rho(x)$. Recall that $K_x(y)=\frac{1}{|x-y|^a}$. Thus, by the Hardy-Littlewood-Sobolev inequality
\begin{equation*}
\int_{\R^3}\rho(x)\,\Nrm{d\Gamma_l(K_x)\,\psi_2}{\mathcal{G}}^2dx\leq C\iint_{\R^6}\frac{\rho(x)\,g(y)}{|x-y|^{2a}}dx\,dy\leq C\Nrm{\rho}{L^{3/(3-2a)}}\Nrm{g}{L^1}, 
\end{equation*}
where $$g(y)=\Nrm{\psi_2^{(n,m)}(y,x_1,\dots,x_{n-1},y_1,\dots,y_m)}{L^2(dx_1\dots dx_{n-1}\,dy_1\dots dy_m)}^2,$$ being $\psi_2^{(n,m)}$ the $(n,m)$-sector of $\psi_2$ in $\mathcal{G}$. Notice that the last estimate entails the restriction $a<\frac{1}{2}$.
\smallskip

{\bf Step 5. Auxiliary fluctuation dynamics.} We are left with the terms in $\widetilde{P}$ that do not commute with $\mathcal{N}$ nor present cancellations due to the commutator structure. To deal with them, we modify the generator of the fluctuation dynamics $\mathcal{U}(t,s)$ using a perturbative argument. 
More precisely, we split the generator $G_t$ into two parts
\begin{equation*}
G_t=\widetilde{G}_t+B_t,
\end{equation*}
where $B_t$ is  small for $N$ large, and $\widetilde{G}_t$ defines a new auxiliary fluctuation dynamics $\widetilde{\mathcal{U}}(t,s)$ as the solution of the Cauchy problem
\begin{equation}\label{eq:auxiliary-dynamics}
i\hbar\partial_t\,\widetilde{\mathcal{U}}(t,s)=\widetilde{G}_t\,\widetilde{\mathcal{U}}(t,s)\,,\quad\quad \widetilde{\mathcal{U}}(s,s)=1,
\end{equation}
whose well-posedness has been shown in \cite[Appendix A]{Chong-Lafleche-Saffirio_2021}.\\
The terms in $\widetilde{P}$ are therefore absorbed into the new generator $\widetilde{G}_t$ and the smallness of $B_t$ allows us to use $\widetilde{G}_t$ instead of $G_t$, paying the price of an additional small error term. This enables us to prove Proposition~\ref{prop:bound-calN}, that together with the estimate in \eqref{eq:densities-number-particles} concludes the proof of the convergence rate for the mean-field approximation from the many-body dynamics \eqref{eq:LvN} to the Hartree-Fock equation \eqref{eq:HF}.


\section{The Vlasov equation}\label{sec:4}

Once obtained the mean-field approximation of the many-body evolution by the solution to the Hartree-Fock equation, it is legitimate to investigate the limit $\hbar\to 0$ in order to get the Vlasov equation. To this end, we Weyl quantize the Vlasov equation \eqref{eq:Vlasov}
\begin{equation}\label{eq:weyl-vlasov}
i\hbar\partial_t\op_f=\com{-\hbar^2\Delta,\op_f}+A_{\op_f},
\end{equation}
where $\op_f$ is the Weyl transform of the solution $f$ to the Vlasov equation and $A_{\op_f}$ denotes the operator with integral kernel
\begin{equation*}
A_{\op_f}(x,y)=\nabla\(K*\rho_f\)\(\frac{x+y}{2}\)\cdot(x-y)\,\op_f(x,y).
\end{equation*}
We are now in the position of comparing $\op_f$ with $\op$, solution to the Hartree-Fock equation \eqref{eq:HF}. \\
In the same spirit of the Bogoliubov transformation in the mean-field context (see Step 2 in Section~\ref{sec:MF}), we define the unitary transformation $\mathcal{U}(t,s)$ as the two-parameter semigroup, solution to the Cauchy problem
\begin{equation*}
i\hbar\partial_t\,\mathcal{U}(t,s)=H_{\op}(t)\,\mathcal{U}(t,s),\quad\quad \mathcal{U}(s,s)=1,
\end{equation*}
where $H_{\op}(t)$ is the time-dependent Hartree-Fock Hamiltionian defined in \eqref{eq:HF-hamiltonian}. The semigroup $\mathcal{U}(t,s)$ plays a similar role to the one of the Bogoliubov transformation, namely it changes the reference frame entailing some cancellations.
More precisely, by conjugating the difference $(\op-\op_f)$ with respect to $\mathcal{U}(t,s)$, the contributions given by the kinetic part of \eqref{eq:HF-hamiltonian} and the right-hand side of \eqref{eq:weyl-vlasov} disappear, leading to  
\begin{equation*}
\begin{split}
i\hbar\partial_t\,\mathcal{U}^*(t,s)(\op-\op_f)\,\mathcal{U}(t,s)=&\ \mathcal{U}^*(t,s)\com{K*(\rho-\rho_f),\op_f}\,\mathcal{U}(t,s)\\
&+\mathcal{U}^*(t,s)B(\op_f)\,\mathcal{U}(t,s)\\ &+\mathcal{U}^*(t,s)\com{X_{\op},(\op-\op_f)}\,\mathcal{U}(t,s)
\end{split}
\end{equation*}
with $B(\op_f)$ the operator with integral kernel
\begin{equation*}
\begin{split}
B&(\op_f)(x,y)\\ &\ =\left[\(K*\rho_f\)(x)-\(K*\rho_f\)(y)-\nabla\(K*\rho_f\)\(\frac{x+y}{2}\)\cdot(x-y)\right]\op_f(x,y).
\end{split}
\end{equation*}
By Duhamel's formula and taking the trace norm, we get
\begin{equation}\label{eq:duhamel}
\begin{split}
\Nrm{\op-\op_f}{\L^1}\leq&\Nrm{\op^\init-\op_f^\init}{\L^1}+\frac{1}{\hbar}\int_0^t \Nrm{\com{K*(\rho-\rho_f),\op_f}}{\L^1}ds\\
&+\frac{1}{\hbar}\int_0^t \Nrm{B(\op_f)}{\L^1}ds+\frac{1}{\hbar}\int_0^t \Nrm{\com{X_{\op},(\op-\op_f)}}{\L^1}ds,
\end{split}
\end{equation}
where we used that $\mathcal{U}(t,s)$ is a unitary operator. We now estimate each term on the right-hand side of \eqref{eq:duhamel}.\smallskip

\subsection{Error terms} The term $B(\op_f)$ as well as the exchange term $X_{\op}$ turn out to be sub-leading in the cases we are interested in, namely $a\in(0,1]$. It has been proven in \cite[Proposition~4.4]{Lafleche-Saffirio_2020} that 
\begin{equation}\label{eq:B-L1}
\Nrm{B(\op_f)}{\L^1}\leq C\hbar^2\,\Nrm{\rho_f}{L^1\cap H^m}\Nrm{\nabla_v^2}{H^{2n}_{2n}(\R^6)},
\end{equation}
where $m=(n+a-1)$ and $n>\frac{3}{2}$, and $H_{2n}^{2n}(\R^6)$ denotes the Hilbert space $W^{2n,2}(\R^6)$ weighted with $\(1+|x|^2+|v|^2\)^{n}$. Taking into account the factor $\hbar^{-1}$ in the second line of \eqref{eq:duhamel}, we conclude that the term containing $B(\op_f)$ gives a contribution of order $\hbar$. As for the term containing the exchange operator, we rely on \cite[Proposition~5.1]{Lafleche-Saffirio_2020}, that proves the following bound:
\begin{equation}\label{eq:exchange-L1}
\Nrm{\com{X_{\op},(\op-\op_f)}}{\L^1}\leq c\hbar^{3-a}\,\Nrm{|\opp|^{\frac{a}{2}}\op}{\L^2}\(\Nrm{\op}{\L^1}+\Nrm{\op_f}{\L^1}\).
\end{equation}
Taking into account the factor $\hbar^{-1}$ in the second line of \eqref{eq:duhamel} we conclude that the term containing $X_{\op}$ gives a contribution of order at most $\hbar$ because, if $a\in(0,1]$, $\hbar^{-1}\hbar^{3-a}\leq \hbar$ for $\hbar\ll 1$. Therefore, for the class of interaction potentials we are considering, the exchange term does not change the order of the rate of convergence given by the term $B(\op_f)$. 
\smallskip

\subsection{Leading order term} The main contribution comes from the commutator term $\com{K*(\rho-\rho_f),\op_f}$. Indeed, writing explicitly the convolution we obtain
\begin{equation}\label{eq:convolution}
\Nrm{\com{K*(\rho-\rho_f),\op_f}}{\L^1}\leq \int |\rho
(x)-\rho_f(x)|\,\Nrm{\com{K(x-\cdot),\op_f}}{\L^1}dx. 
\end{equation}
To cancel the factor $\hbar^{-1}$ in front of the time integral in the second term of the first line of the right-hand side of \eqref{eq:duhamel} we seek for some smallness arising from the commutator structure. More precisely, the following estimate holds true.
\begin{proposition}[Theorem~4.1 in \cite{Lafleche-Saffirio_2020}]
\label{prop:commutator-est} Let $\fb=\frac{3}{a+1}$ and $\fb'$ be the conjugated H\"{o}lder exponent of $\fb$. Then for $\eps>0$ and $\tilde{\eps}\in\(0,\frac{\eps}{2\fb'}\)$, there exists $C>0$ such that
\begin{equation*}
\Nrm{\com{K(x-\cdot),\op_f}}{\L^1}\leq C\hbar\,\Nrm{\diag(|\Dhv\op_f|)}{L^{\fb'-\eps}}^{\frac{1}{2}+\tilde{\eps}}\Nrm{\diag(|\Dhv\op_f|)}{L^{\fb'+\eps}}^{\frac{1}{2}-\tilde{\eps}}.
\end{equation*}
\end{proposition}
Notice that Proposition~\ref{prop:commutator-est} provides a uniform bound in the $x$ variable on the trace norm of the commutator $\com{K(x-\cdot),\op_f}$, hence the integral in $x$ on the right-hand side of \eqref{eq:convolution} is bounded by the $L^1$-norm of the difference of the spatial densities $\rho$ and $\rho_f$. By duality and using that $\rho=\diag(\op)$, $\rho_f=\diag(\op_f)$, we obtain the bound
\begin{equation*}
\Nrm{\rho-\rho_f}{L^1}=\sup_{J\in L^\infty,\ \Nrm{J}{L^\infty}\leq 1}\left|\int_{\R^3} J(x)\,\(\rho(x)-\rho_f(x)\)\,dx\right|\leq \Nrm{\op-\op_f}{\L^1},
\end{equation*}
that allows as to close the the Gr\"{o}nwall-type inequality. More precisely we get
\begin{equation*}
\Nrm{\op-\op_f}{\L^1}\leq \(\Nrm{\op^\init-\op_f^\init}{\L^1}+C_0(t)\hbar+C_1(t)\hbar^{2-a}\)e^{\lambda(t)},
\end{equation*}
with $C_1(t),\,C_2(t)$ and $\lambda(t)$ functions depending only on weighted Sobolev norms of the solution to the Vlasov equation, for which the regularity theory is well-established (see for instance \cite{Lions-Perthame_1991,Pfaffelmoser_1992} and \cite[Appendix~A]{Lafleche-Saffirio_2020}).


\section{Conclusions and open problems}\label{sec:5}

Despite the recent progresses, the analysis on time intervals of order one of the most interesting case of particles interacting via the Coulomb potential remains a major open problem, as does the companion problem of deriving the Vlasov equation with Coulomb interaction from the dynamics of many classical particles. 
In the context of classical mechanics, the derivation problem can be formulated as follows. We consider a $N$-particle configuration on the phase space $(x_1,v_1,x_2,v_2,\dots,x_N,v_N)\in\R^{6N}$. Its evolution in time is given by the Newton equations 
\begin{equation}
\frac{d\,x_i}{dt}(t)=v_i(t),\quad\quad \frac{d\,v_i}{dt}(t)=-\frac{1}{N}\sum_{j=1}^N\nabla V(x_i(t)-x_j(t)), \quad\quad i=1,\dots,N,
\end{equation}
where $V:\R^3\to\R$ is a two-body interaction potential.
The problem of justifying the Vlasov equation~\eqref{eq:Vlasov} starting from the dynamics of $N$ particles obeying Newton's laws has been proved for smooth potentials in the pioneering works \cite{Neunzert_1972, Braun-Hepp_1977, Dobrushin_1979} (see also~\cite{Spohn_1991}). The class of potentials was then extended to locally H\"{o}lder continuous interactions in \cite{Hauray-Jabin_2007, Hauray-Jabin_2015}. In \cite{Boers_2015} the convergence towards the Vlasov equation is proven for potentials with a vanishing cut-off (as $N\to\infty$) converging to singular interactions, including the Coulomb potential. A further improvement has been achieved in \cite{Lazarovici-Pickl_2017}, where the size of the cut-off is comparable to the mean inter-particle distance. Moreover, in \cite{Grass_2021} a class of potentials slightly singular at zero has been treated. More recently, Serfaty \cite{Serfaty} provided a proof of the derivation in the Coulomb case for the special class of initial data called monokinetic. Thus the derivation of the Vlasov equation in the cases of Coulomb and gravitational interactions, which are the relevant models for applications to plasma physics and astrophysics, is still an open problem.


\vspace{0.5cm} \indent {\it
A\,c\,k\,n\,o\,w\,l\,e\,d\,g\,m\,e\,n\,t\,s.\;} The authors
acknowledge support by the NCCR SwissMAP and the Swiss National Science Foundation through the Eccellenza project PCEFP2\_181153.

\bigskip
\begin{center}

\end{center}

\bigskip
\bigskip
\begin{minipage}[t]{10cm}
\begin{flushleft}
\small{
\textsc{Chiara Saffirio}
\\*University of Basel,
\\*Department of Mathematics and Computer Science
\\*Spiegelgasse 1
\\* Basel, 4051, Switzerland
\\*e-mail: chiara.saffirio@unibas.ch 
}
\end{flushleft}
\end{minipage}

\end{document}